\documentclass[10pt]{article}

\usepackage{OSAmeet09_arxiv}
\usepackage{amsmath}
\usepackage{amstext}
\usepackage{amsxtra}
\usepackage{amssymb}
\usepackage{graphicx}
\usepackage{subfigure}

\newcommand {\rhovec}{\ensuremath \boldsymbol{\rho}}

\begin{document}
\addtocounter{footnote}{1}
\title{Defeating Passive Eavesdropping with\\ Quantum Illumination}
\author{Jeffrey H. Shapiro}
\address{Research Laboratory of Electronics, Massachusetts Institute of Technology, Cambridge, MA 02139, USA}
\email{jhs@mit.edu}

\begin{abstract}
Quantum illumination permits Alice and Bob to communicate at 50\,Mbit/s over 50\,km of low-loss fiber with error probability less than $10^{-6}$ while the optimum passive eavesdropper's error probability must exceed 0.28.   
\end{abstract}


We introduce a new optical communication protocol, based on quantum illumination \cite{Secure}, for defeating passive eavesdropping.  The communication system functions as follows.  Alice uses spontaneous parametric downconversion (SPDC) to produce $M$ signal-idler mode pairs, with annihilation operators $\{\,\hat{a}_{S_m}, \hat{a}_{I_m} : 1 \le m \le M\,\}$, whose joint density operator is the tensor product of independent, identically distributed (iid) zero-mean, Gaussian states for each mode pair with the common Wigner-distribution covariance matrix 
\begin{equation}
{\boldsymbol \Lambda}_{SI} = \frac{1}{4}\left[\begin{array}{cccc}
S  & 0 & C_q & 0 \\ 
0 & S  & 0 & -C_q \\
C_q & 0 & S & 0 \\ 
0 & -C_q & 0 & S
\end{array}\right], \label{quadent}
\end{equation}
where $S \equiv 2N_S + 1$ and $C_q \equiv 2\sqrt{N_S(N_S+1)}$, and $N_S$ is the average photon number of each signal (and idler) mode.  Alice sends her signal modes to Bob, over a pure-loss channel, retaining her idler modes.  Bob receives modes with annihilation operators $\hat{a}_{B_m} = \sqrt{\kappa}\,\hat{a}_{S_m} + \sqrt{1-\kappa}\,\hat{e}_{B_m}$,  where the $\{\hat{e}_{B_m}\}$, are in their vacuum states.  Bob  imposes an  identical, binary phase-shift keyed (BPSK) information bit ($k = 0$ or 1 equally likely) on each $\hat{a}_{B_m}$.  He then employs a phase-insensitive optical amplifier with gain $G$, and transmits the amplified modulated modes, $\hat{a}_{B_m}' \equiv (-1)^k\sqrt{G}\,\hat{a}_{B_m} + \sqrt{G-1}\,\hat{a}^\dagger_{N_m}$, back to Alice through the same pure-loss channel, where the $\{\hat{a}_{N_m}\}$ are in iid thermal states with $\langle\hat{a}_{N_m}\hat{a}^\dagger_{N_m}\rangle = N_B/(G-1) \ge 1$.  Alice receives modes with annihilation operators $\hat{a}_{R_m} = \sqrt{\kappa}\,\hat{a}_{B_m}' + \sqrt{1-\kappa}\,\hat{e}_{A_m}$, where the $\{\hat{e}_{A_m}\}$ are in their vacuum states.  Given Bob's information bit $k$, we have that $\rhovec_{RI}^{(k)}$, the joint state of Alice's $\{\hat{a}_{R_m},\hat{a}_{I_m}\}$ modes, is the tensor product of iid, zero-mean,  Gaussian states for each mode pair with the common Wigner covariance matrix
\begin{equation}
{\boldsymbol \Lambda}^{(k)}_{RI} = \frac{1}{4} \left[\begin{array}{cccc}
A  & 0 & (-1)^kC_a & 0 \\ 
0 & A  & 0 & (-1)^{k+1}C_a \\
(-1)^kC_a & 0 & S & 0 \\ 
0 & (-1)^{k+1}C_a & 0 & S
\end{array}\right]\!\!, \label{quadentrcv}
\end{equation}
where $A \equiv 2\kappa^2 G N_S + 2\kappa N_B + 1$ and $C_a \equiv \kappa \sqrt{G}\,C_q$.  

Eve is a passive eavesdropper who collects \em all\/\rm\ the photons that are lost en route from Alice to Bob and from Bob to Alice, i.e., she observes $\hat{c}_{S_m} = \sqrt{1-\kappa}\,\hat{a}_{S_m} - \sqrt{\kappa}\,\hat{e}_{B_m}$ and 
$\hat{c}_{R_m} = \sqrt{1-\kappa}\,\hat{a}_{B_m}' -\sqrt{\kappa}\,\hat{e}_{A_m}$ for $1\le m\le M$.
Given Bob's bit value, Eve's joint density operator, $\rhovec_{c_Sc_R}^{(k)}$, is the tensor product of $M$ iid mode-pair density operators that are zero-mean, jointly Gaussian states with Wigner covariance matrix
\begin{equation}
{\boldsymbol \Lambda}^{(k)}_{c_Sc_R} 
= \frac{1}{4}
 \left[\begin{array}{cccc}
D  & 0 & (-1)^kC_e & 0 \\ 
0 & D  & 0 & (-1)^kC_e \\
(-1)^kC_e & 0 & E & 0 \\ 
0 & (-1)^kC_e & 0 & E
\end{array}\right]\!\!, \label{quadenteve}
\end{equation}
where $D \equiv 2(1-\kappa)N_S + 1$, $C_e \equiv 2(1-\kappa)\sqrt{\kappa G}\,N_S$, and $E \equiv 2(1-\kappa)\kappa G N_S + 2(1-\kappa)N_B + 1$.  

Exact error probabilities for these Gaussian-state hypothesis tests are not easy to evaluate, so we shall rely on quantum Chernoff bounds, which we can calculate using the results from \cite{Pirandola}.  In Fig.~1 we compare the Chernoff bounds for Alice and Eve's optimum quantum receivers for a particular case, along with an error-probability lower bound on Eve's optimum quantum receiver.  Alice's error probability \em upper\/\rm\ bound can be orders of magnitude lower than the Eve's error probability \em lower\/\rm\ bound when both use optimum quantum reception despite Eve's getting the lion's share of the photons.  Moreover, using an algebraic computation program we have found the following approximate forms for the Chernoff bounds on Alice and Eve's optimum quantum receivers:
$\Pr(e)_{\rm Alice} \le \exp(-4M\kappa G N_S/N_B)/2$ and 
$\Pr(e)_{\rm Eve} \le \exp(-4M\kappa(1-\kappa)G N_S^2/N_B)/2$,
which apply in the low-brightness, high-noise regime, viz., when $N_S \ll 1$ and $\kappa N_B \gg 1$.  They imply that Alice's Chernoff bound error exponent will be orders of magnitude \em higher\/\rm\ than that of Eve in this regime, and so the advantageous quantum-illumination behavior shown in Fig.~1 is typical for this regime.
\begin{figure}[h]
\begin{center}
\includegraphics[width=2.25in]{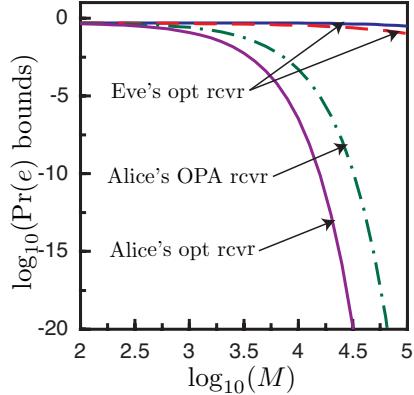}
\end{center}
\vspace*{-.2in}
\caption{Error-probability bounds for $N_S = 0.004$, $\kappa=0.1$, $G = N_B = 10^4$.  Solid curves: Chernoff bounds for Alice and Eve's optimum quantum receivers.  Dashed curve: error-probability lower bound for Eve's optimum quantum receiver.  Dot-dashed curve: Bhattacharyya bound for Alice's OPA receiver.}
\end{figure}

While we will accord Eve the right to an optimum quantum receiver,  let us show that Alice can still enjoy an enormous advantage in error probability when she uses  a version of Guha's optical parametric amplifier (OPA) receiver  \cite{Guha}, i.e., a receiver we know how to build.  Here Alice uses an OPA to obtain modes given by $\hat{a}'_m \equiv \sqrt{G_{\rm opa}}\,\hat{a}_{I_m} + \sqrt{G_{\rm opa}-1}\,\hat{a}_{R_m}^\dagger,$ where $G_{\rm opa} = 1 + N_S/\sqrt{\kappa N_B}$, and then makes her bit decision based on the photon-counting measurement $\sum_{m=1}^M\hat{a}'^\dagger_m\hat{a}'_m$.  The Bhattacharyya bound on this receiver's error probability in the $N_S \ll 1$, $\kappa N_B \gg 1$ regime turns out to be
$\Pr(e)_{\rm OPA} \le \exp(-2M\kappa G N_S/N_B)/2$,
which is only 3\,dB inferior, in error exponent, to Alice's optimum quantum receiver.  We have included the numerically-evaluated Bhattacharyya bound for Alice's OPA receiver in Fig.~1.    

Two final points deserve note.  BPSK communication is phase sensitive, so Alice's receiver will require phase coherence that must be established through a tracking system.  More importantly, there is the path-length versus bit-rate tradeoff.  Operation must occur in the low-brightness regime.  So, as channel loss increases, Alice must increase her mode-pair number $M$ at constant $N_S$ and $G$ to maintain a sufficiently low error probability \em and\/\rm\ communication security.  For a $T$-sec-long bit interval and $W$\,Hz SPDC phase-matching bandwidth, $M = WT$  implies that her bit rate will go down as loss increases at constant error probability.  With $W = 1$\,THz and $T = 20\,$ns, so that $M = 2\times 10^4$,  the case shown in Fig.~1 will yield 50\,Mbit/s communication with $\Pr(e)_{\rm OPA} \le 5.09 \times 10^{-7}$ and $0.285 \le \Pr(e)_{\rm Eve} \le 0.451$ when Alice and Bob are linked by 50\,km of  0.2\,dB/km loss fiber, assuming that the rest of their equipment is ideal.  

In conclusion, we have shown that quantum illumination can provide immunity to passive eavesdropping in a lossy, noisy environment despite that environment's destroying the entanglement produced by the source.

This work was supported by the Office of Naval Research Basic Research Challenge Program, the W. M. Keck Foundation for Extreme Quantum Information Theory, and the DARPA Quantum Sensors Program.

\end{document}